\documentclass[twocolumn,showpacs,showkeys,preprintnumbers,amsmath,amssymb]{revtex4}

\usepackage{graphicx}
\usepackage{dcolumn}
\usepackage{bm}

\begin{document}

\title{Dynamic fluctuations in the superconductivity of NbN films from 
	microwave conductivity measurements}

\author{Takeyoshi Ohashi}
\author{Haruhisa Kitano}
\author{Astutaka Maeda}
 \affiliation{
	Department of Basic Science, The University of Tokyo,\\
	3-8-1, Komaba, Meguro-ku, Tokyo 153-8902, Japan
 }

\author{Hiroyuki Akaike}
\author{Akira Fujimaki}
 \affiliation{
	Department of Quantum Engineering, Nagoya University,\\
	Furocho, Chikusa-ku, Nagoya-shi, Aichi, 464-8603, Japan}

\date{\today}

\begin{abstract}
We have measured the frequency and temperature dependences of 
complex ac conductivity, 
$\sigma(\omega)=\sigma_1(\omega)-i\sigma_2(\omega)$, 
of NbN films in zero magnetic field between 0.1 to 10 GHz 
using a microwave broadband technique.
In the vicinity of superconducting critical temperature, $T_c$, 
both $\sigma_1(\omega)$ and $\sigma_2(\omega)$ 
showed a rapid increase 
in the low frequency limit owing to the fluctuation effect of superconductivity.
For the films thinner than 300 nm, 
frequency and temperature dependences of fluctuation conductivity, 
$\sigma_{\textrm{fl}}(\omega,T)$, 
were successfully scaled onto one scaling function, 
which was consistent with the Aslamazov and Larkin model 
for two dimensional (2D) cases.
For thicker films, 
$\sigma_{\textrm{fl}}(\omega,T)$ data could not be scaled, 
but indicated that 
the dimensional crossover from three dimensions (3D) to 2D 
occurred 
as the temperature approached $T_c$ from above.
This provides a good reference of 
ac fluctuation conductivity 
for more exotic superconductors of current interest.
\end{abstract}



\maketitle

\section{\label{sec:1}Introduction}

The effects of thermal fluctuation near the superconducting phase transition have  been 
an interesting issue for many years, 
both theoretically and experimentally \cite{Tinkham}.
These effects attract much attention particularly for 
strongly type-II superconductors of recent interest, 
for example, 
the high-$T_c$ cuprate superconductors, 
the organic superconductors, 
and the heavy Fermion superconductors,
because the effects of thermal fluctuation are enhanced 
by the short coherence length and the anisotropy \cite{FFH}.

The superconducting fluctuation leads to 
a divergence 
of a number of physical quantities, 
such as, the specific heat, the susceptibility, 
the dc nonlinear electrical conductivity,
and the ac linear electrical conductivity.
One of the remarkable features of the superconducting fluctuation 
is a scaling behavior appearing in these diverging quantities.
Detailed analyses of such scaling behavior provide important information 
on the universality class of the second order phase transition.
Among these physical quantities, ac conductivity is 
one of the most powerful probes to investigate the fluctuation effect, 
because the frequency and temperature dependences of complex conductivity 
enable the direct verification of a dynamic scaling theory \cite{Halperin}.
Although the theory predicts the scaling behavior 
in other dc quantities,
as well as in the frequency dependence of the ac conductivity,
$\sigma(\omega)$, 
the latter has an additional advantage.
The scaling analysis generally needs two scaling parameters to be determined.
It is, however, impossible to determine these two parameters uniquely 
from the dc (not complex) experimental data.
Therefore, the scaling analyses of dc data usually assume 
a particular dimension 
in order to establish a certain relationship between these two parameters. 
This reduces the degrees of freedom in the scaling analyses 
from two to one, leading to less convincing analyzed results.
Contrary to this, complex $\sigma(\omega)$
can provide the unique determination of the scaling parameters 
without any assumptions.

Recently, the dynamic scaling analyses on ac conductivity have been performed 
for some kinds of high-$T_c$ cuprates.
Booth {\it et al.} \cite{Booth} investigated 
optimally doped YBa$_2$Cu$_3$O$_{7-\delta}$ thin films 
by microwave measurement 
and observed a critical fluctuation 
with a large dynamic critical exponent, $z\sim 2.3-3.0$.
Corson {\it et al.} \cite{Corson} investigated 
underdoped Bi$_2$Sr$_2$CaCu$_2$O$_{8+\delta}$ thin films 
by THz techniques 
and observed thermally generated vortices, 
which are explained in the 2D-{\it XY} model.
In spite of these studies, 
a general consensus on the fluctuation of high-$T_c$ cuprates
has not yet been achieved.
This is, in part, due to the lack of systematic studies on high-$T_c$ cuprates 
and fundamental studies on conventional superconductors 
which are very important as reference results 
for many more exotic superconductors of recent interest.
Previous studies on the ac fluctuation conductivity 
of conventional superconductors have been published. 
For instance, 
Lehoczky and Briscoe \cite{Lehoczky} 
measured the temperature dependence of fluctuation conductivity 
on lead films 
at several fixed microwave frequencies.
Tanner \cite{Tanner} measured the temperature dependence 
of the characteristic frequency of the fluctuation on lead films 
by far-infrared transmission.
Although these studies were in agreement with the calculation on 
Gaussian theory \cite{Schmidt}, 
the frequency dependence was insufficient to examine the scaling analysis.
In addition, 
these studies compared the measured conductivity or 
some other physical quantities 
with the theoretical prediction based on a particular microscopic model.
In contrast, 
dynamic scaling analysis works without any explicit calculations 
of the individual models. 
This is the most significant merit of dynamic scaling. 
However, there are no studies of dynamic scaling analysis of ac fluctuation conductivity 
on conventional superconductors in zero magnetic field,
though this approach has been applied to the vortex-glass transition in conventional superconductor, indium \cite{Okuma}.

In this paper, 
we report the frequency and temperature dependences of 
ac conductivity of films of a conventional superconductor, NbN, 
and perform dynamic scaling analysis 
on excess conductivity due to the superconducting fluctuation effects in zero magnetic field.
The comprehensive examinations of the fluctuation conductivity, 
including the scaling behavior itself, critical exponents, and also
the explicit temperature and frequency dependences 
of fluctuation conductivity, 
clarify the validity and the limit of the dynamic scaling analysis.

\section{Summary of Theoretical background}

The first successful theory of the dc fluctuation conductivity 
in superconductors 
was provided by Aslamazov and Larkin \cite{AL}.
They considered the thermal relaxation of the order parameter to 
calculate the direct contribution of the superconducting pairs 
created by fluctuations (AL-term), as follows,
\begin{eqnarray}
\sigma_{\textrm{dc}}^{\textrm{2DAL}}
	&=&\frac{1}{16}\frac{e^2}{\hslash t}\varepsilon^{-1},
							\nonumber \\
\sigma_{\textrm{dc}}^{\textrm{3DAL}}
	&=&\frac{1}{32}\frac{e^2}{\hslash \xi_0}\varepsilon^{-1/2},
							\label{eq:ALDC}
\end{eqnarray}
where $t$ is the thickness, $\xi_0$ is the coherence length at $0$ K,
and $\varepsilon=|T/T_c-1|$.
These predictions were in good agreement with experimental results 
on dirty superconductors \cite{Glovoer}.
Maki \cite{MT-Maki} and Thompson \cite{MT-Thompson}
proposed the existence of an additional term (MT-term) 
owing to an indirect contribution of the superconducting fluctuation
on the quasiparticle conductivity, as follows,
\begin{eqnarray}
\sigma_{\textrm{dc}}^{\textrm{2DMT}}
	&=&\frac{1}{8}\frac{e^2}{\hslash t}
	\frac{1}{\varepsilon-\delta}
	\ln\left(\frac{\varepsilon}{\delta}\right),
							\nonumber \\
\sigma_{\textrm{dc}}^{\textrm{3DMT}}
	&=&\frac{1}{8}\frac{e^2}{\hslash \xi_0}\varepsilon^{-1/2},
							\label{eq:MTDC}
\end{eqnarray}
where $\delta$ is the pair-breaking parameter introduced to avoid 
an unphysical divergence of conductivity at $T>T_c$ in the 2D case.
The MT term explained a larger magnitude and an anomalous 
temperature dependence of the fluctuation conductivity 
observed on cleaner superconductors \cite{Strongin,MaskerParks}.

As for the dynamic aspect of the fluctuation, 
the frequency dependence of the AL-term was calculated 
by Schmidt \cite{Schmidt} 
using the time-dependent Ginzburg-Landau equation, as follows,
\begin{widetext}
\begin{eqnarray}
\sigma^{\textrm{2DAL}}(\omega)
	&=& \sigma_{\textrm{dc}}^{\textrm{2DAL}}
	S^{\textrm{2DAL}}
	\left(\frac{\pi\hslash\omega}{16k_B T_c\varepsilon}\right),
							\nonumber \\
\textrm{Re} S^{\textrm{2DAL}}(x)
	&=& \frac{2}{x}\tan^{-1}{x}-\frac{1}{x^2}\ln(1+x^2),
							\nonumber \\
\textrm{Im} S^{\textrm{2DAL}}(x)
	&=& \frac{2}{x^2}(\tan^{-1}{x}-x)+\frac{1}{x}\ln (1+x^{2}),
							\nonumber \\		\sigma^{\textrm{3DAL}}(\omega)
	&=& \sigma_{\textrm{dc}}^{\textrm{3DAL}}
	S^{\textrm{3DAL}}
	\left(\frac{\pi\hslash\omega}{16k_B T_c\varepsilon}\right),
							\nonumber \\
\textrm{Re} S^{\textrm{3DAL}}(x)
	&=& \frac{8}{3x^2}
		\left [
		1-(1+x^2)^{3/4}\cos\left(\frac{3}{2}\tan^{-1}{x}\right)
		\right ] ,
							\nonumber \\
\textrm{Im} S^{\textrm{3DAL}}(x)
	&=& \frac{8}{3x^2}
		\left [
		-\frac{3}{2}x
		+(1+x^2)^{3/4}\sin\left(\frac{3}{2}\tan^{-1}{x}\right)
		\right ] .
							\label{eq:ALac}
\end{eqnarray}
\end{widetext}
On the other hand, the frequency dependence of the MT-term 
was calculated by Aslamazov and Varlamov \cite{MTac}.
Starting from a layered superconductor 
they found, 
in the 2D and 3D limits, 
that the contribution of the MT-term should be added to the AL-term, 
as follows,
\begin{widetext}
\begin{eqnarray}
\textrm{Re} S^{\textrm{2DAL+MT}}(x) 
	&=& \textrm{Re} S^{\textrm{2DAL}}(x) 
		+ \frac{2\pi x-2\ln{2x}}{1+4x^2},
								\nonumber \\
\textrm{Im} S^{\textrm{2DAL+MT}}(x) 
	&=& \textrm{Im} S^{\textrm{2DAL}}(x)
		+ \frac{\pi+4x\ln{2x}}{1+4x^2}, 
								\nonumber \\
\textrm{Re} S^{\textrm{3DAL+MT}}(x) 
	&=& \textrm{Re} S^{\textrm{3DAL}}(x)
		+\frac{4-4x^{1/2}+8x^{3/2}}{1+4x^2},
								\nonumber \\
\textrm{Im} S^{\textrm{3DAL+MT}}(x) 
	&=& \textrm{Im} S^{\textrm{3DAL}}(x)
		+\frac{4x^{1/2}-8x+8x^{3/2}}{1+4x^2}.
							\label{eq:MTac}
\end{eqnarray}
\end{widetext}

The appearance of the high-$T_c$ cuprate superconductors enabled us 
to recognize importance of the critical fluctuation even in the 
physics of superconductivity.
It is expected that higher $T_c$, shorter coherence length,
longer magnetic penetration depth, 
and quasi-two dimensionality enhance 
the temperature region 
where the Gaussian theory is not valid.
To study such a critical region,
Fisher, Fisher and Huse \cite{FFH} formulated a dynamic scaling rule 
on fluctuation conductivity 
in the vicinity of a second order phase transition, 
as follows,
\begin{equation}
\sigma_{\textrm{fl}}(\omega) 
	= \sigma_0 S\left( \frac{\omega}{\omega_0}\right),
							\label{eq:scaling}
\end{equation}
where $S$ is a complex universal scaling function,
and $\sigma_0$, $\omega_0$ are scaling parameters 
whose temperature dependences are related to that of 
a correlation length, $\xi$, which diverges at $T_c$,
as follows,
\begin{eqnarray}
\sigma_0(\varepsilon) &\propto & [\xi(\varepsilon)]^{z+2-d},
							\nonumber \\
\omega_0(\varepsilon) &\propto & [\xi(\varepsilon)]^{-z},
							\nonumber \\
\xi(\varepsilon) &=& \xi_0\varepsilon^{-\nu},
							\label{eq:para}
\end{eqnarray}
where $\nu$ is a static critical exponent, 
$z$ is a dynamic critical exponent, and 
$d$ is an effective spatial dimension.
In the limit of $\omega/\omega_0 \to \infty$,
$\sigma_{\textrm{fl}}=|\sigma_{\textrm{fl}}|e^{i\phi_{\textrm{fl}}}$
behaves as
\begin{eqnarray}
|\sigma_{\textrm{fl}}|(\omega) 
	&\propto& \omega^{-(z+2-d)/z},
							\nonumber\\
\phi_{\textrm{fl}}(\omega) 
	&=& \frac{\pi}{2}\frac{z+2-d}{z}.
							\label{eq:FFH_lim}
\end{eqnarray}
Thus, keeping these formulas in mind,
$\nu$, $z$, and $d$ can be obtained by the  
$\sigma_{\textrm{fl}}(\omega, T)$ measurement, and the universality class
of the transition can be determined.

\begin{figure}[t]
\includegraphics[width=0.8\linewidth]{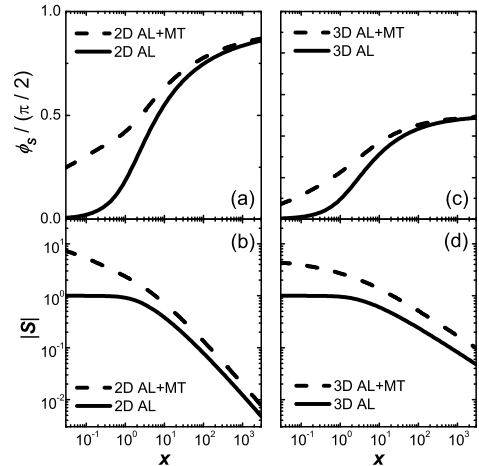}
\caption{\label{fig:S} Phase, $\phi_S(x)$, and amplitude $|S(x)|$ of theoretical scaling functions for the AL term and the AL term + the MT term for (a),(b) the 2D and (c),(d) the 3D case.}
\end{figure}

Note that this scaling theory is a general one 
which also contains the Gaussian result as a special case; 
$z=2, \nu=0.5$ \cite{Dorsey}.
In this case, the scaling functions are exactly the same as given
in Eqs. (\ref{eq:ALac}) and (\ref{eq:MTac}).
The Gaussian results of the scaling functions are shown in Fig \ref{fig:S}.

\section{\label{sec:2}Experimental}

\begin{table*}
\caption{
\label{tab:sample}
Various parameters of the NbN films; $t$ : thickness, $\rho$ : dc resistivity, $R_\square$ : sheet resistance, $T_c^R$ : the resistive $T_c$, $T_c^{\textrm{MF}}$ : the mean-field $T_c$, $T_c^{\textrm{scale}}$ : the $T_c$ used for the scaling analysis, $\xi_0$ : the coherence length at 0 K, $l$ : the electronic mean free path, and $\varepsilon_c$ : the crossover temperature from 3D to 2D. See the text for details.}
\begin{ruledtabular}
\begin{tabular}{ccccccccc}
$t$ (nm)& $\rho$(m$\Omega$cm) &$R_\square$($\Omega$)& $T_c^R$ (K)&$T_c^{\textrm{MF}}$(K) &$T_c^{\textrm{scale}}$(K)& $\xi_0$ (nm)& $l$ (nm) & $\varepsilon_c$\\
\hline
50 &	0.27&	54&	12.81&	12.87&	12.865&	3.6& 0.20 &	$5.1\times 10^{-2}$\\
150&	0.20&	20&	14.24&	14.36&	14.360&	3.9& 0.27 &	$6.7\times 10^{-3}$\\
300&	0.19&	6.3&	14.89&	14.90&	-&	4.1& 0.29 &	$1.8\times 10^{-3}$\\
450&	0.22&	4.9&	15.09&	15.12&	-&	4.2& 0.25 &	$8.6\times 10^{-4}$\\
\end{tabular}
\end{ruledtabular}
\end{table*}
\subsection{\label{sec:2-1}Fabrication and characterization of NbN films}

NbN films were deposited on LaSrAlO$_4$(LSAO) (001) 
substrate by a reactive sputtering technique, 
details of which were described elsewhere \cite{NbNfilm}.
The thicknesses of prepared films, $t$'s, 
were 50 nm, 150 nm, 300 nm, and 450 nm, respectively.
NbN was a suitable reference for
high-$T_c$ cuprates because 
its short coherence length (2 - 4 nm) and 
long penetration depth (200 - 600 nm) \cite{Mathur,Saito,Shoji}, 
are relatively similar to those of high-$T_c$ cuprates.
We selected LSAO substrates for the direct comparison 
of NbN data with those of La$_{2-x}$Sr$_{x}$CuO$_4$ epitaxial thin films on LSAO \cite{Kitano}.

\begin{figure}[b]
\includegraphics[width=\linewidth]{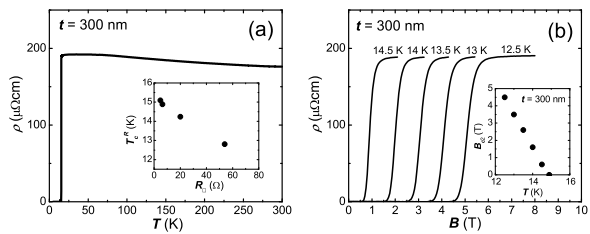}
\caption{\label{fig:NbNdc}(a) Temperature dependence of dc resistivity, $\rho$, of the NbN film with $t$ = 300 nm. Inset: Sheet resistance dependence of $T_c^R$. (b) Magnetic field dependence of dc resistivity, $\rho$, of the NbN film with $t$ = 300 nm. Inset: Temperature dependence of $B_{c2}$.}
\end{figure}

Figure \ref{fig:NbNdc}(a)
shows the dc resistivity, $\rho$, of the film with 300 nm
measured by an ordinary four-probe method.
The resistive critical temperatures, $T_c^R$, 
were defined as the temperature at which dc resistivity vanishes,
and the electronic mean free paths, $l$, were estimated using the relationship, 
$l = 9\times10^{11}\hbar(3\pi^2)^{1/3}[e^2\rho(n^{2/3}\times0.6)]^{-1}$,
where we used the carrier concentration $n \approx 2.4\times10^{23}$ cm$^{-3}$ 
and the ratio of the area of the Fermi surface to that of a free-electoron Fermi surface
$\sim 0.6$ \cite{Mathur}.
They are listed in Table \ref{tab:sample}, together with some important parameters.
As shown in the inset of Fig. \ref{fig:NbNdc}(a), 
$T_c^R$ decreased linearly with the sheet resistance, $R_\square$.
This sheet resistance dependence was consistent with 
a prediction by a theory of weak localization \cite{localization},
indicating that a series of these films 
was sufficiently uniform to discuss the thickness dependence.

As shown in Fig. \ref{fig:NbNdc}(b),
we also measured 
dc resistivity under dc magnetic fields applied parallel to the film 
in order to estimate the upper critical filed, $B_{c2}$.
Using the relationships $B_{c2}(0) \approx 0.7 dB_{c2}(T)/dT |_{T=T_c} T_c$ 
\cite{WWH} and 
$\xi_0=(\Phi_0/2\pi B_{c2}(0))^{1/2}$, 
where $\Phi_0$ is the flux quantum, 
$\xi_0$ was evaluated, which is also listed in Table \ref{tab:sample}.
All the films satisfied the condition, $l\ll\xi_0$, guaranteeing that they are in the dirty limit.

\subsection{\label{sec:2-2}Microwave broadband method for ac conductivity measurement}

We measured the complex reflection coefficient, $S_{11}$,
of the film placed at the end of the transmission line.
Since this is a non-resonant method
that is free from the frequency restriction of resonant methods,
the detailed frequency dependence can be obtained.
The complex impedance, $Z_L$, and the complex conductivity, $\sigma$, 
of the sample can be obtained 
from $S_{11}$, as follows \cite{BoothRSI},
\begin{eqnarray}
	Z_L&=&\frac{1+S_{11}}{1-S_{11}}Z_0,	\nonumber \\
	Z_L&=&\sqrt{\frac{i\mu\omega}{\sigma}}
	\coth(\sqrt{i\mu\omega\sigma}t),	\label{eq:sigma}
\end{eqnarray}
where $Z_0 = 377$ $\Omega$ is the impedance of free space, 
and $\mu$ is a permeability, which can be approximated by 
that of vacuum in this paper.
When $t$ is much less than the skin depth of the sample 
($\sim$ 10 $\mu\textrm{m}$ for NbN at 10 GHz), 
Eq. (\ref{eq:sigma}) can be approximated
 as follows,
\begin{equation}
	Z_L = \frac{1}{t\sigma}.\label{eq:sigma2}
\end{equation}
Therefore, $\sigma$ can be derived directly from $S_{11}$ 
for sufficiently thin films.
This is one reason why we used films in this study.

Equation (\ref{eq:sigma}) shows 
that even a small error in $S_{11}$ 
can have a large effect on the calculation of $\sigma$, 
particularly when $Z_L$ is much smaller than $Z_0$.
In practice, a condition, $|Z_L| >$ 0.2 $\Omega$, is required 
because the error in $|S_{11}|$ is about 0.01 dB 
and $Z_L$ = 0.2 $\Omega$ corresponds to $|S_{11}| = -0.01$ dB.
The typical normal resistivity of NbN is 200 $\mu\Omega\textrm{cm}$.
Thus, only films thinner than 1 $\mu\textrm{m}$ 
satisfy this condition, 
if we tend to measure $\sigma(\omega)$ in the vicinity of $T_c$,
where $|Z_L|$ is 10 times lower than in the normal state.

\subsection{\label{sec:2-3}Measurement setup}

The microwave broadband method is a two-probed measurement
affected by the contact resistance.
To reduce the contact resistance, 
gold electrodes 200 nm thick were sputtered 
on the surface of the films 
and annealed in air at 250 $^\circ$C for 1 hour.
The electrodes in the samples for microwave measurement 
were made into the so-called Corbino-disk geometry.
The diameters of the inner and outer electrodes were
1.0 mm and 2.4 mm, respectively.

The film was connected to a coaxial cable (RG 405/U) 
through a modified 2.4 mm jack-to-jack coaxial adapter
(M/A-COM OS-2.4 adaptor 8580-0000-02).
A spring-loaded gold pin in the center conductor of the modified adapter 
and a spring set behind the substrate
made a stable electrical contact
between the sample and the transmission line
even at low temperatures \cite{BoothRSI,KitanoISS}.
The other end of the transmission line was connected to a vector network 
analyzer (HP8510C) to measure $S_{11}$ 
over the frequency range from 45 MHz to 50 GHz. 
The network analyzer was operated in the step-sweep mode 
to ensure the phase coherence at each frequency. 
The inner and outer conductors of the coaxial cable were made of Cu 
except for the last 15 cm in the neighborhood of the sample, 
where CuNi was used to reduce the thermal flow into the sample.
The length of the whole transmission line was $\sim$ 1.7 m 
and its loss was $\sim$ 5 dB at 10 GHz.
Incident microwave power was $0$ dBm, so that 
the amplitude of the current density in the film was 
$\sim 1000$ Acm$^{-2}$.
$S_{11}$ did not show the incident power dependence 
at this power range for $T>T_c$ indicating that the linear conductivity was measured.
We also measured dc resistance, $R$, 
just before and after the measurement of $S_{11}(\omega)$, 
by connecting a dc current source and a voltmeter 
to the transmission line 
through the bias port of the network analyzer.

All measurements were performed 
after keeping the sample holder in the cryostat at least for 10 hours 
in order to ensure the equilibrium of temperature distribution 
in the transmission line.
During each measurement, the temperature was fixed to a constant value
within $\pm$1 mK.

\subsection{\label{sec:2-4}Calibration}

The experimentally measured reflection coefficient includes 
the extrinsic attenuation, reflection, and phase shift 
due to the transmission line {\it etc}. 
This problem becomes more complicated for measurements 
at low temperatures, 
because the conductivity and the length of the transmission line 
considerably vary with temperature.

To calibrate, we performed the following procedure
at each frequency and temperature.
The  measured reflection coefficient, 
$S_{11}^{\textrm{m}}$, can be expressed as follows,
\begin{equation}
S_{11}^{\textrm{m}}=E_D+\frac{E_RS_{11}}{1-E_SS_{11}},\label{eq:cal}
\end{equation}
where $E_D$, $E_R$, and $E_S$ are complex error coefficients 
representing 
the directivity, the reflection tracking, and the source mismatch, 
respectively \cite{calib}.
Equation (\ref{eq:cal}) implies that
a set of three independent measurements using loads 
with known $Z_L$ (or $S_{11}$)
are needed to determine three error coefficients \cite{Stutzmann}.
Once they are obtained, 
all systematic errors 
can be eliminated from $S_{11}^{\textrm{m}}$ data, 
utilizing Eq. (\ref{eq:cal}).
However, all errors are not completely systematic nor reproducible,
because there are small differences in the experimental configurations 
among different measurement runs for the sample and the known loads.
This becomes serious, especially 
in measurements of samples with small impedance.
To overcome this difficulty, 
we regard the $|S_{11}|(\omega)$ of the sample in the superconducting state 
at 10 K (the lowest temperature of the measurement) 
to be equal to that of an ideal short ($Z_L(\omega)=0$),
and $S_{11}(\omega)$ of the normal state at 20-25 K 
(the temperatures sufficiently far from the fluctuation region) 
to that of an ideal load ($Z_L(\omega)=R(\omega=0)$).
This assumption is reasonable 
because (1) the expected loss of NbN at 10 K ($\sim 10^{-5}$ dB) 
is much less than the resolution of our measurement, 
(2) the Drude relaxation rate of NbN ($\sim 10^{14} $ s$^{-1}$ \cite{tau}) is 
much higher than the measured frequency, 
and (3) the change in the property of the transmission line is negligible 
in the range of temperature concerned.
With this procedure, 
the error in the calibrated $S_{11}$ was reduced 
to 0.01 dB in magnitude, 
and 0.05$^\circ$ in the phase at 1 GHz.

\section{Results and Discussion}
\begin{figure}
\includegraphics[width=\linewidth]{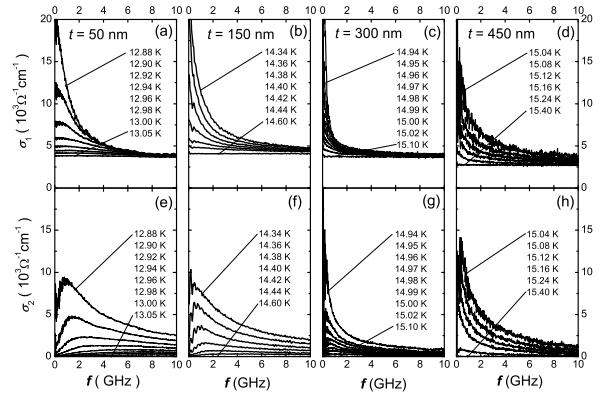}
\caption{\label{fig:sigma}
Frequency dependence of (a)-(d) $\sigma_1$ and (e)-(h) $\sigma_2$
at several temperatures near $T_c$}
\end{figure}
Figure \ref{fig:sigma} shows the frequency dependence 
of complex conductivity of NbN films 
at several temperatures near $T_c$.
As the temperature approaches $T_c$ from above, 
both the real part, $\sigma_1$, and the imaginary part, $\sigma_2$, 
showed a tendency to diverge in the low frequency limit.
In addition, a characteristic frequency of $\sigma(\omega)$ decreased, 
suggesting that 
the contribution of the superconducting fluctuation 
to $\sigma(\omega)$ was evident and 
the relaxation time of the fluctuation became longer.
\begin{figure}
\includegraphics[width=\linewidth]{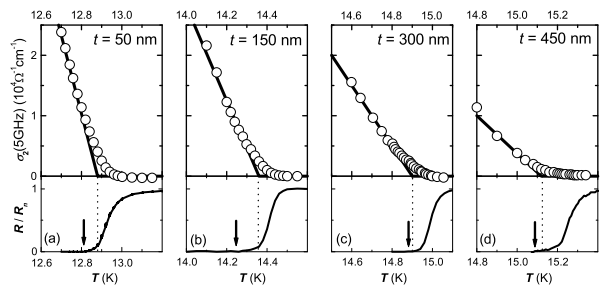}
\caption{\label{fig:sigma2}
Temperature dependence of $\sigma_2$ at 5 GHz (open circles) and $R$ normalized by the normal resistance, $R_n$ (solid lines in the lower panels). The bold solid lines in the upper panels are the mean-field fitting. The dotted-lines and the arrows indicate $T_c^{\textrm{MF}}$ and $T_c^R$, respectively.} 
\end{figure}

Figure \ref{fig:sigma2} shows the temperature dependence 
of $\sigma_2$ at 5 GHz, together with the dc resistance, $R$.
Within a mean-field approximation, 
$\sigma_2$ is proportional to the superfluid density, and should vary as 
$\sigma_2(T) \propto (1-T/T_c)$ just below $T_c$, 
whereas $\sigma_2(T) = 0$ above $T_c$, 
as were shown by the bold straight lines in Fig. \ref{fig:sigma2}. 
The experimental data clearly show that 
the discontinuous superconducting transition changes
to a continuous one interconnected by 
a fluctuation-dominated region.
This intermediate region corresponded to 
the rounding of $R(T)$. In this case, 
the transition temperature, $T_c^{\textrm{MF}}$, 
was determined by the fitting to a mean-field theory, 
and almost agreed with $T_c^{R}$.
Rather, $T_c^{\textrm{MF}}$ was 
a better definition of $T_c$ than $T_c^{R}$, 
since dc resistivity contains 
the fluctuation-induced finite resistance even below $T_c$,
and is sensitive to a local decrease of $T_c$ owing to, for example, 
the thickness distribution.

\begin{figure}
\includegraphics[width=\linewidth]{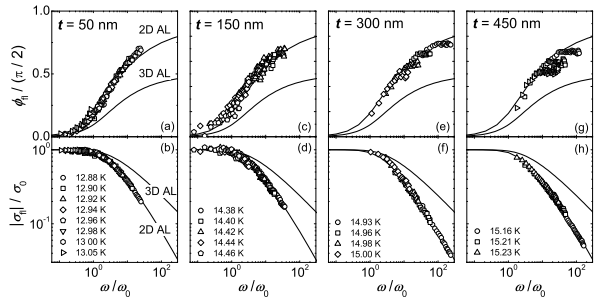}
\caption{\label{fig:scale}
The results of the dynamic scaling analysis 
on the phase, $\phi_{\textrm{fl}}$, and the magnitude, $|\sigma_{\textrm{fl}}|$ 
for the NbN films 
with $t$ = 50 nm ((a),(b)), 150 nm ((c),(d)), 300 nm ((e), (f)), and 450 nm((g), (h)).
The attempts for the films with $t$ = 300 nm and 450 nm were in vain.
Here we used the data at 0.4 - 6, 0.2 - 9 GHz, 0.3 - 3.5 GHz, and 0.3 - 3.5 GHz 
for the film with $t$ = 50 nm, 150 nm, 300 nm, and 450 nm, respectively.
The solid lines represent theoretical calculations for 2D-AL and 3D-AL (Eq. (\ref{eq:ALac})).
Note that the scaling parameters, $\omega_0$ and $\sigma_0$ were derived from these plots.}
\end{figure}
Next, we analyzed the fluctuation conductivity, $\sigma_{\textrm{fl}}$, 
in detail. 
We subtracted the mean-field conductivity 
from the total measured conductivity 
to extract the fluctuation contribution.
Here we focused on the fluctuation at $T>T_c$,
thus the mean-field conductivity is the normal conductivity, $\sigma_n$.
Therefore,
\begin{equation}
\sigma_{\textrm{fl}}(\omega,T) \equiv \sigma(\omega,T)-\sigma_n(\omega).
\end{equation}
Figure \ref{fig:scale} 
shows the result of the scaling of the amplitude, $|\sigma_{\textrm{fl}}|$, 
and the phase, $\phi_{\textrm{fl}}$.
For the films with $t=50$ and $150$ nm,
the scaled data were in good accordance with 
the theoretically calculated $S(\omega/\omega_0)$ 
in the 2D-AL model without the MT-term (Eq. (\ref{eq:ALac})).
The absence of the MT-term contribution is reasonable for the dirty films.
Contrary to this, the scaling analysis 
failed for the films with $t=300$ and $450$ nm.

\begin{figure}
\includegraphics[width=0.9\linewidth]{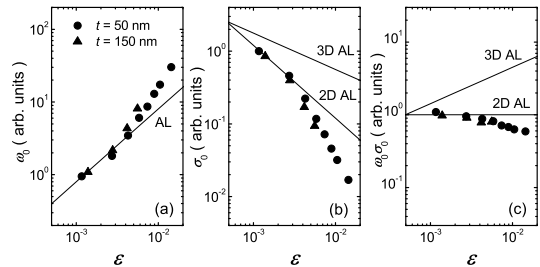}
\caption{\label{fig:para}
Temperature dependence of (a) $\omega_0$, (b) $\sigma_0$, 
and (c) $\omega_0\sigma_0$ , where $\varepsilon = T/T_c-1$.
The solid lines represent theoretical calculation for the 2D-AL model ($\nu=1/2, z=2, d=2$) and the 3D-AL model ($\nu=1/2, z=2, d=3$).}
\end{figure}
The temperature dependence of the scaling parameters,
$\omega_0$ and $\sigma_0$, 
also confirmed that the observed fluctuation was the 2D-AL fluctuation.
Before the quantitative estimation 
of the critical exponents,
it should be noted that the product $\omega_0\sigma_0$
had little temperature dependence (Fig. \ref{fig:para}(c)).
This strongly suggested $d=2$ 
according to Eq. (\ref{eq:para}).
To obtain the critical exponent, $\nu$ and $z$, 
$T_c$ has to be determined and $\omega_0$ and $\sigma_0$ have to be plotted 
as a function of $\varepsilon = T/T_c-1$.
We determined $T_c^{\textrm{scale}}$ to obtain the best fit of $\omega_0(\varepsilon)$ and 
$\sigma_0(\varepsilon)$ to the theoretical prediction.
As was also shown in Table \ref{tab:sample}, 
the obtained $T_c^{\textrm{scale}}$'s were in fairly good agreement with $T_c^{\textrm{MF}}$'s.
As shown in Fig. \ref{fig:para},
$\omega_0(\varepsilon)$ and $\sigma_0(\varepsilon)$ 
are in good agreement with the 2D-AL fluctuation, 
in which $\nu=1/2$, $z=2$, and $d=2$, 
rather than the 3D-AL fluctuation ($\nu=1/2$, $z=2$, and $d=3$).
The deviation from the theoretical value at higher temperatures is
possibly due to the short-wavelength cutoff effects \cite{Cutoff}.
Here, we should keep in mind 
that the ambiguity in 
the choice of $T_c^{\textrm{scale}}$ largely 
affects the critical exponents.
Thus, it is essential to crosscheck if 
the determined $T_c$, the critical exponents, 
and the dimension are consistent with each other
in all measurements and analyses.
In this study, we obtained the fully consistent data set
of $R(T)$, $\sigma_2(T)$, $S(\omega/\omega_0)$, 
$\omega_0(\varepsilon)$, and $\sigma_0(\varepsilon)$, 
which yields the identical $T_c$, the critical exponents, and the dimension.

\begin{figure}
\includegraphics[width=0.8\linewidth]{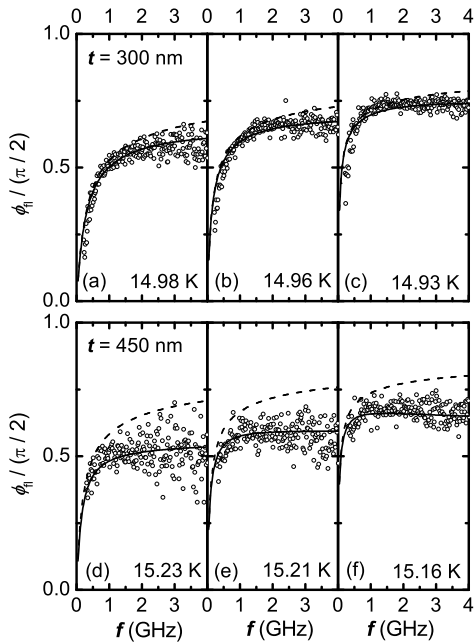}
\caption{\label{fig:Dcross}
Frequency dependence of $\phi_{\textrm{fl}}$ of
NbN films with (a)-(c) $t=300$ nm, and (d)-(f) $t=450$ nm.
The dash lines are the pure 2D calculation (Eq. (\ref{eq:ALac})).
The solid lines are the calculation for the case of the dimensional crossover 
(Eq. (\ref{eq:series})).}
\end{figure}
The observed 2D behavior was apparently puzzling
because 3D-AL fluctuation was expected 
for a three dimensional isotropic superconductor like NbN.
We believe that this decrease in the dimensionality 
stemmed from the size effect. 
In fact, when $\xi(T) \gg t$,
the effective spatial dimension would become 2D.
The failure of the dynamic scaling in the thicker films (Fig. \ref{fig:scale}(e)-(g))
also can be explained by the dimensional crossover due to the size effect 
because the scaling law is valid only for the limit of 2D or 3D.
The effective spatial dimension would change from 3D to 2D
as $\xi(T)$ exceeds $t$ when temperatures approaches $T_c$.

To discuss this size effect more quantitatively,
we refer to the calculation of the fluctuation conductivity of 
thin films with finite thickness by Schmidt \cite{Schmidt}.
By introducing discrete components of the wave vector 
normal to the plane of the film, satisfying the boundary condition 
of vanishing derivative at the surface, 
the fluctuation conductivity was obtained as follows,
\begin{eqnarray}
\label{eq:series}
\sigma(\omega)&=&\sum_{\nu}\frac{1}{1+q_\nu^2\xi^2}
\> \sigma^{\textrm{2DAL}}
\left( \frac{\omega}{1+q_\nu^2\xi^2}\right)\\
&&q_{\nu} = \nu\frac{\pi}{t}	\qquad \nu=0,1,2\cdots.\nonumber
\end{eqnarray}
For $\varepsilon \ll \varepsilon_c$ ($\xi \gg t$) only the first term of the series in Eq. (\ref{eq:series}) 
gives an appreciable contribution, which is reduced to Eq. (3);
contribution from other terms entering for $\varepsilon \sim \varepsilon_c$ or 
even $\varepsilon > \varepsilon_c$ will spoil the 2DAL-scaling approach.
As shown in Fig. \ref{fig:Dcross}, $|\phi_{\textrm{fl}}|(\omega)$ 
of the films with $t=300$ and $450$ nm, 
can be described by Eq. (\ref{eq:series}) rather than by the pure 2D formula (Eq. (\ref{eq:ALac})).
This indicates that the films cannot be regarded as 2D any longer, 
suggesting the breakdown of the dynamic scaling behavior.

Although there is no definite criteria to separate the 2D region from the 3D region, 
we defined a characteristic temperature, 
$\varepsilon_{\textrm{c}}=(\pi\xi_0/t)^2$, 
by equating $\xi=t/\pi$, below which the pure 2D picture would be expected.
In Table \ref{tab:sample}, the estimated values of 
$\varepsilon_{\textrm{c}}$ are listed.
For the films with $t=50$ and $150$ nm, 
the experimentally observed 2D-AL behavior at temperatures 
below $\varepsilon_{\textrm{c}}$ is reasonable.
On the other hand, for the films with $t=300$ and $450$ nm, 
the measurement temperatures were in the crossover region from 2D to 3D.
In that region, 
the frequency dependence of $\sigma_{\textrm{fl}}$ largely 
changes within the measurement temperature range, 
so that the data did not collapse to a unique scaling function.
Consequently, the dynamic scaling analysis is inappropriate.

\section{conclusion}
We measured the frequency and temperature dependences 
of microwave conductivity of NbN films using a microwave broadband method.
We observed the superconducting-fluctuation-induced excess conductivity
in the vicinity of $T_c$.
For the films with $t = 50$ and $150$ nm, 
dynamic scaling analysis on fluctuation conductivity yielded 
2D-AL behavior.
The temperature dependence of $R$ and $\sigma_2$, 
the obtained scaling function, 
and the critical exponents  were fully consistent.
On the other hand,
for the films with $t = 300$ and $450$ nm, 
whose sample dimensions were comparable to $\xi(T)$ 
in the measurement temperature range, 
the dynamic scaling theory failed to explain the data 
because of the dimensional crossover.
This was an experimental evidence for the breakdown of the scaling behavior
owing to the dimensional crossover.
These data both validate the dynamic scaling theory and clarify its limits, 
and serve as a reference for the 
application of scaling analysis for more exotic superconductors 
of current interest.

\section*{ACKNOWLEDGEMENTS}
We thank K. Gomez for useful comments on the manuscript.
This work was partly supported by 
the Grant-in-Aid for Scientific Research 
(13750005 and 15760003) 
from the Ministry of Education, Culture, Sports, Science and Technology of Japan.
T. Ohashi thanks 
the Japan Society for the Promotion of Science for financial support.

\newpage 

%
\end{document}